\documentclass[twocolumn,aps,showpacs,superscriptaddress]{revtex4}
\usepackage{amssymb}
\usepackage{amsmath}
\usepackage{amsfonts}
\usepackage{amsmath,amssymb}
\usepackage{graphicx}
\usepackage{bm}
\usepackage{multirow}

\newcommand{\vect}{\boldsymbol}
\begin{document}

\title{Moir\'e minibands in graphene heterostructures with almost\\ commensurate $\sqrt3\times\sqrt3$ hexagonal crystals}

\author{J.~R.~Wallbank}
\affiliation{Department of Physics, Lancaster University, Lancaster, LA1 4YB, UK}

\author{M.~Mucha-Kruczy\'{n}ski}
\affiliation{Department of Physics, Lancaster University, Lancaster, LA1 4YB, UK}
\affiliation{Department of Physics, University of Bath, Claverton Down, Bath, BA2 7AY, United Kingdom}

\author{V.~I.~Fal'ko}
\affiliation{Department of Physics, Lancaster University, Lancaster, LA1 4YB, UK}

\date{\today}

\begin{abstract} 
We present a phenomenological theory of the low energy moir\'e minibands of Dirac electrons in graphene placed on an almost commensurate hexagonal underlay with a unit cell approximately three times larger than that of graphene.
A slight incommensurability results in a periodically modulated intervalley scattering for electrons in graphene.
In contrast to the perfectly commensurate Kekul\'e distortion of graphene, such supperlattice perturbation leaves the zero energy Dirac cones intact,
but is able to open a band gap at the edge of the first moir\'e subbband, asymmetrically in the conduction and valence bands.
\end{abstract}


\maketitle 
%%%%%%%%%%%%%%%%% Introduction %%%%%%%%%%%%%%%%%
Two alternative methods exist to create long-period superlattices for two-dimensional (2D) electrons. One method, developed for semiconductors, is based on the lithographic patterning of the semiconductor surface \cite{semiconductor_superlattices}. 
The other method, highlighted by the studies of 2D atomic crystals,  arises naturally from the existence of quasi-periodic moir\'e patterns formed by two slightly incommensurate 2D lattices with similar crystal symmetry, placed on top of each other. Graphene on hexagonal boron nitride is one example of such heterostructure, where the effect of the moir\'e  superlattice on 2D electrons leads to  pronounced changes in the electronic properties detected by STM \cite{marchini_prb_2007,diaye_njp_2008,decker_nanolett_2011}, and magnetotransport experiments \cite{ponomarenko_nature_2013,dean_nature_2013,hunt_science_2013}.

The specific form of moir\'e superlattice for graphene electrons, generated by a hexagonal underlay, depends on the ratio between the periods of the two lattices and their mutual orientation. 
The abundance of layered hexagonal crystals and semiconductors with a hexagonal surface layer, allows for a multiplicity of qualitatively different superlattice structures, with various levels of moir\'e super-cell complexity. The simplest and, by now, best studied is the highly orientated graphene-hBN  heterostructure.
Here we analyze the second simplest moir\'e pattern for Dirac electrons in graphene produced by a hexagonal underlay with an elementary unit cell approximately 3 times bigger than that of graphene. The effect of a perfectly commensurate $\sqrt 3\times\sqrt3$ superlattice, known as the Kekul\'e distortion of the honeycomb lattice \cite{cheianov_ssc_2009}, consists in the Bragg type intervalley scattering of graphene electrons, which opens a gap between the conduction and valence bands. A hexagonal underlay with the lattice constant $a_S=\sqrt 3 (1+\delta)a$, $|\delta|\ll1$, slightly different from that of the Kekul\'e superlattice of graphene and a small misaligned angle $\theta$, produce a periodically oscillating intervalley coupling. Although this does not open a gap in graphene's Dirac point, it creates a specific miniband spectrum, whose generic features are studied in this paper. Below, we employ a phenomenological approach to classify the possible structure of moir\'e 
minibands  
of Dirac electrons in graphene  \cite{wallbank_prb_12} and, in particular, the behavior of the edge of the first minibands on the conduction and valence band sides. 

%%%%%%%%%%%%%%%%% Fig 1 %%%%%%%%%%%%%%%%%%%%%%%%%%%
\begin{figure}[t]
  \includegraphics[width=0.48\textwidth]{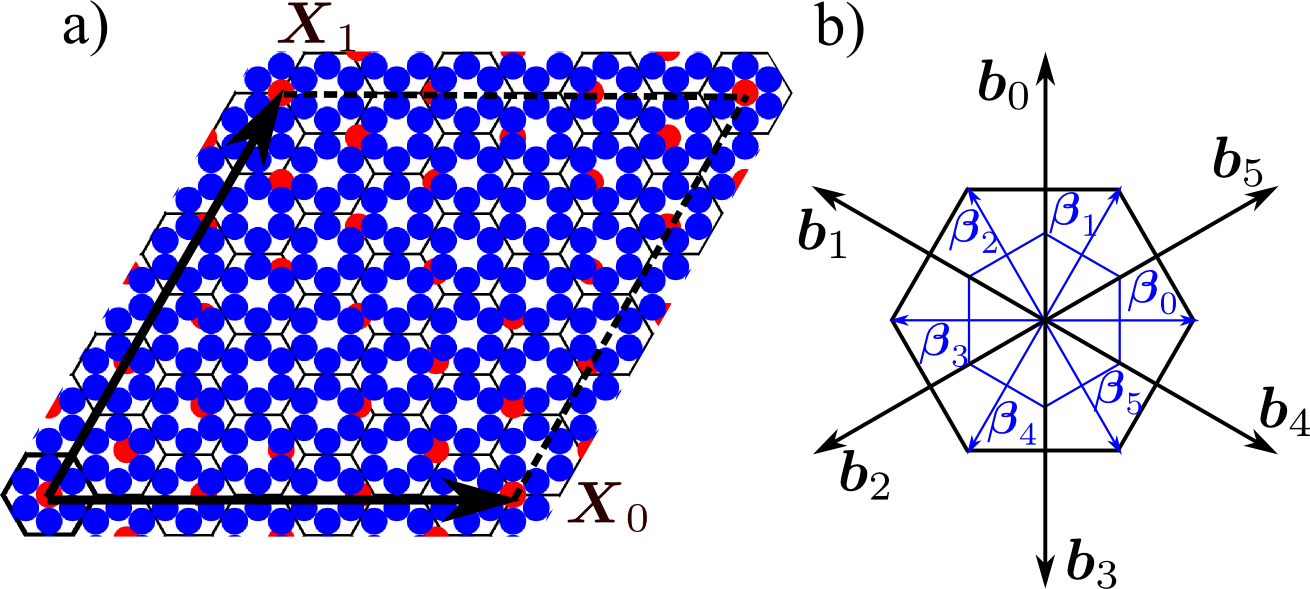}
  \caption{(a) The moire pattern formed from graphene (blue) on a underlay (red) with $\theta=0$, $\delta=\frac{1}{9}$. The black hexagons follow  Kekul\'e lattice of graphene. (b) The two sets of reciprocal lattice vectors, $\vect b_m$ and $\vect \beta_m$, with their associated Brillouin zones.}
  \label{fig:moire_BZ}
 \end{figure}
 
The image of a moir\'e supperlattice for graphene on a substrate with a period almost commensurate with the $\sqrt3\times\sqrt3$ Kekul\'e lattice of graphene is shown in Fig.~\ref{fig:moire_BZ}. Since  graphene electrons belong to the Bloch states in its hexagonal Brillouin zone corners and a Kekul\'e perturbation leads to their intervalley Bragg scattering, the symmetry of the electronic system is described by the group of wavevectors $K_\pm$, equivalent to the extended point group $C_{6v}+ tC_{6v}+t^2C_{6v}$ where $t$ is an elementary translation of the honeycomb lattice. 
That is why in Fig.~\ref{fig:moire_BZ} (a) we show both the actual positions of carbon atoms in graphene and, using lines, the Kekul\'e lattice.
The periodic occurrence of sites from the underlay under equivalent positions of graphene honeycomb lattice is described by a moir\'e pattern which is periodic under translations by $X_0$ and $X_1$. The associated reciprocal lattice vectors belong to the set 
$\check{\vect  b}=\{ \vect b_m=\hat R_{2\pi m/6} \vect b_0\}_{m=0,\cdots,5}$ where $\hat R_\psi$ is the rotation matrix, and $\vect b_0=\left[1-(1+\delta)^{-1}\hat R_\theta\right]\left(0,\frac{4\pi}{\sqrt3 a}\right)$, so that  $|\vect b_n|\equiv b=\frac{4\pi}{\sqrt3 a}\sqrt{\delta^2+\theta^2}$. In contrast, the equivalent positions of substrate sites on the Kekul\'e lattice are characterized by the $\sqrt 3$ times longer period of $X_0+X_1$ and reciprocal lattice vectors from the set $\check{\vect  \beta} =\{\vect \beta_m=\frac{1}{\sqrt3}\hat R_{\frac{-\pi }{2}} \vect b_m\}_{m=0,\cdots,5}$ with    $|\vect \beta_n|\equiv \beta =b/\sqrt3$.
The coexistence of these two periodicities is taken into account, on an equal footing, in the phenomenological Hamiltonian for graphene's Dirac electrons,

%%%%%%%%%%%%%%%%%%%%%%%%%%%%%%% Hamiltonian %%%%%%%%%%%%%%%%%%%%%%%%%%%%%%%%%
\begin{widetext}
  \begin{gather}
 \hat H=v\hat{\vect p} \cdot \vect\sigma 
+U_{E'}v\beta F({\check{\vect \beta}})\sigma_3 
+ U_{G}v  \left[\vect\sigma\times\vect l_z\right]  \cdot\nabla F({\check{\vect \beta}})
+ U_{G'}v\vect\sigma \cdot \nabla F({\check{\vect \beta}}) \label{eq:H}\\
\qquad\qquad \qquad+u_0v b f_1( \check{\vect b})+u_3v b f_2( \check{\vect b})\sigma_3\tau_3+u_1v \left[\vect l_z\times\nabla f_2( \check{\vect b})\right]\cdot\vect \sigma\tau_3+  u_2 v  \tau_3\vect \sigma \cdot\nabla f_2( \check{\vect b}); \nonumber\\
   f_1( \check{\vect v})=\sum_{m=0,\cdots,5}e^{i\vect v_m\cdot\vect r},\qquad 
   f_2( \check{\vect v})=i\sum_{m=0,\cdots,5}(-1)^me^{i\vect v_m\cdot\vect r},\qquad
 F({\check{\vect v}})= f_1( \check{\vect v})\tau_1+f_2( \check{\vect v})\tau_2. \nonumber 
 \end{gather}
 \end{widetext}
This Hamiltonian is written in terms of the Pauli matrices $\sigma_i$ and $\tau_j$ which act separately on the sublattice $(A,B)$ and valley $(K_+, K_-)$ components of the 4-spinors $\left(\psi_{AK_+},\psi_{BK_+},\psi_{BK_-},-\psi_{AK_-}\right)^T$ describing graphene electrons. 
Hence, the second line describes intravalley Bragg scattering, whereas the first line accounts for intervalley scattering.
In writing $\hat H$, we use the earlier observation \cite{yankowitz_natphys_2012,ortix_prb_2012,kindermann_prb_2012,wallbank_prb_12, twisted_blg_lopes_dos_santos,twisted_blg_bistritzer} that the potential felt by the graphene electrons is smoothened by the larger separation between graphene and the substrate than the carbon-carbon distance in graphene. For graphene on hBN, as well as  twisted bilayer graphene, this resulted in the presence of only the simplest set of harmonics, $\check{\vect b}$, in the moir\'e perturbation \cite{yankowitz_natphys_2012,ortix_prb_2012,kindermann_prb_2012,wallbank_prb_12, twisted_blg_lopes_dos_santos,twisted_blg_bistritzer}. For graphene on a almost commensurate $\sqrt 3\times\sqrt3$ hexagonal underlay the same argument leads to the appearance of the intervalley terms. In Eq.~\eqref{eq:H}, the relative strength of moir\'e perturbations, measured in the unit of energy $vb=\sqrt3 v\beta$, is set by dimensionless parameters  $U_{E'}$, $U_
G$, $U_{G'}$, $u_{i=0,1,2,3}$. Here, we assume that such moir\'e perturbation is small, $|U_i|\ll1$, $|u_j|\ll1$, and that the underlay has an inversion-symmetric unit cell, which is a natural approximation \cite{footnote:inversion} for a simple monoatomic surface layer.

%%%%%%%%%%%%%%%%%%%%%%%%%%%  Microscopic Models %%%%%%%%%%%%%%%%%%%%%%%%%%%%%%%%%%%%%
 To supplement a phenomenological approach to describe the moir\'e supperlattice, Eq.~\eqref{eq:H}, we also estimated parameter $U_i$ and $u_j$ for two limiting microscopic models: (a)  the underlay is modeled  as hexagonal lattice of point charges \cite{wallbank_prb_12}, and (b) the underlay is modeled as a lattice of atomic orbitals on to which the graphene electrons can hop (adapted from a model of twisted bilayer graphene \cite{kindermann_prb_2011}). 
Both models produce similar estimates for sets of phenomenological parameters $U_i$ and $u_j$,
\begin{align}
&v\beta \{U_{E'},U_{G},U_{G'}\}= \tilde V\left\{\frac{1}{2}, \frac{- \delta}{\sqrt{\delta^2\!+\!\theta^2}},	  \frac{\theta}{\sqrt{\delta^2\!+\!\theta^2}}\right\},\label{eq:model}  \\
&vb\{u_{0},u_{1},u_{2},u_{3}\}= \tilde v\left\{\frac{1}{2}, \frac{- \delta}{\sqrt{\delta^2\!+\!\theta^2}},	  \frac{\theta}{\sqrt{\delta^2\!+\!\theta^2}},- \frac{\sqrt 3}{2}\right\}. \nonumber
\end{align}
However model (a) predicts $\tilde V\gg\tilde v$, whereas model (b) predicts $\tilde V \sim \tilde v$ \cite{footnote:model}.

The features of the miniband spectrum of the Dirac electrons prescribed by the intravalley terms, $u_j$, in the second line of Eq.~\eqref{eq:H} have already been explored in studies of graphene on hBN  \cite{yankowitz_natphys_2012,ortix_prb_2012,kindermann_prb_2012,wallbank_prb_12}. The characteristic feature, present in the low energy graphene band structure for this case, consist in the formation of additional mini Dirac points \cite{yankowitz_natphys_2012,ortix_prb_2012,wallbank_prb_12} in a gapless spectrum. In contrast, intervalley perturbations $U_i$, are able to open gaps in the spectrum at the edges of the low energy moir\'e minibands.
Hence, we focus on the role of the intervalley terms, and explore the parameter space $\left[ U_{E'}, U_G, U_{G'}\right]$,   classifying the resulting electron spectra. 
It is useful to notice that for the Hamiltonian in Eq.~\eqref{eq:H} 
\begin{align}
-\epsilon_{-\!U_{E'},U_G,U_{G'}}\!(\!\vect k\!)\!=\!\epsilon_{U_{E'},U_G,U_{G'}}\!(\!\vect k\!)\!=\!\epsilon_{-\!U_{E'},\!-\!U_G,\!-\!U_{G'}}\!(\!\vect k\!).\! \label{eq:symm}
\end{align}
The first equality in Eq.~\eqref{eq:symm} allows us to relate the bandstructure of the valence band to that of the conduction band  by flipping the sign of $U_E$. 
Also, it turns out that the parameter $U_{G'}$ affects the miniband spectra of electrons only in the second order, since its first order effect on the electron energies can be removed by the gauge transformation $\vect \psi\rightarrow e^{-i U_{G'} F(\check{ \vect \beta}) }\vect \psi'$. 
 
\begin{figure}[htbp]
\includegraphics[width=0.48\textwidth]{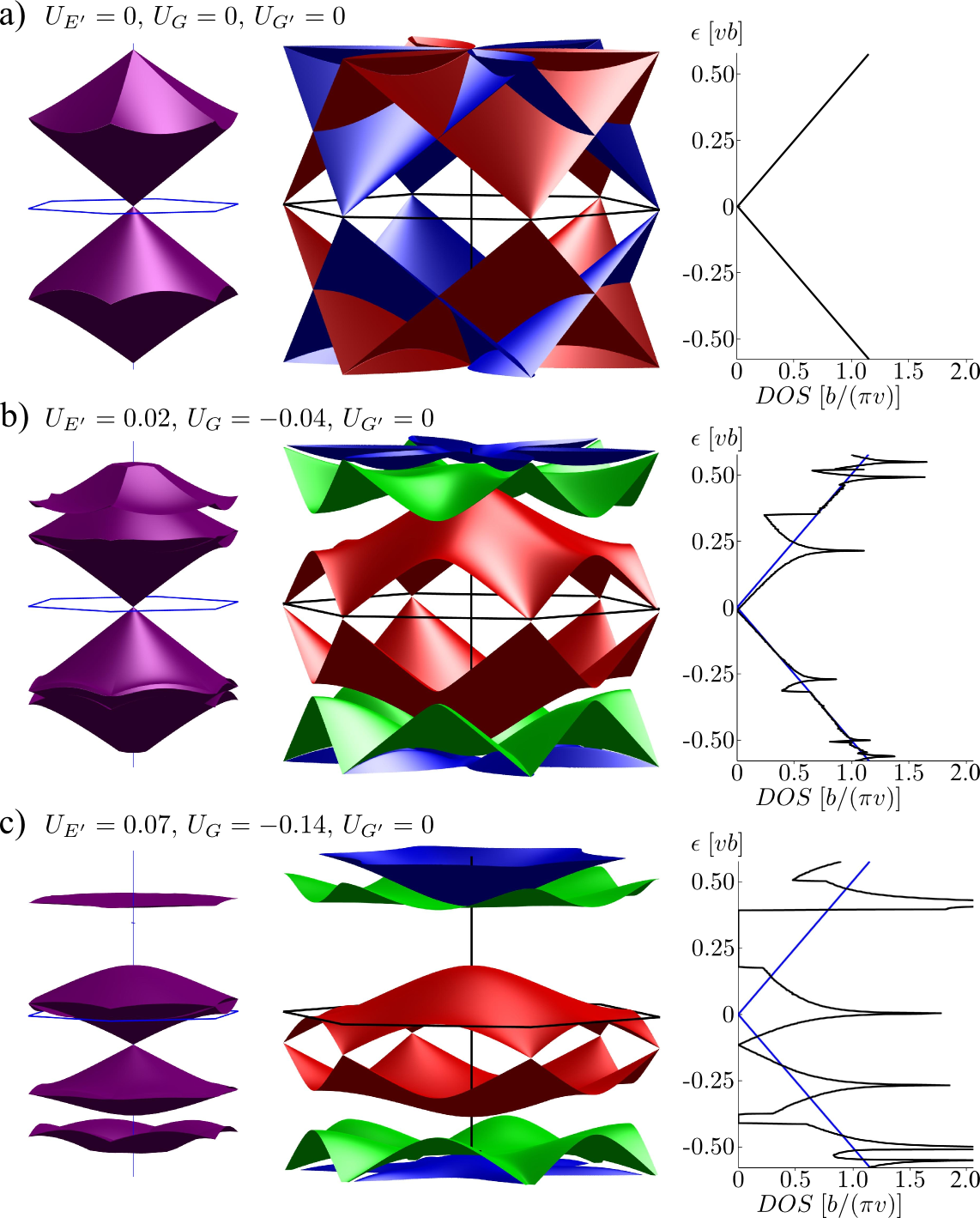}
\caption{Numerically calculated moir\'e minibands shown in the smaller mBZ (left) and larger mBZ (center), and the corresponding density of states (right). 
A Van Hove singularity, originating from the first moir\'e miniband (in both the conduction and valence bands) is always present for the perturbed spectra.
}
\label{fig:bandstructure}
\end{figure}

%%%%%%%%%%%% Translational symmetries %%%%%%%%%
The correspondence between the translational symmetries of the Hamiltonian $\hat H$ and the geometrical symmetry group of the moir\'e supperlattice, $G_{SL}=\{c_6, T_{\vect X_0}\}$, is set by the fact that a translation e.g. by the period $\vect X_0$ indicated in Fig.~\ref{fig:moire_BZ}, is accompanied by a valley-dependent unitary gauge transformation, $\hat U_{t}=-\frac{1}{2}-\frac{\sqrt 3 i}{2}\tau_3$ which represents the effect of the elementary translation of the honeycomb lattice on the 4-component spinors $\vect \psi$. 
This argument establishes the isomorphism of $G_{SL}$ to the symmetry group $G_H=\{\hat c_6,\hat S_{\vect X_0}\}$ of the Hamiltonian $\hat H$, where instead of geometrical translation $T_{\vect X_0}$ we use $\hat S_{\vect X_0}=\hat U_{t}\hat T_{\vect X_0}$ (and $\hat S_{\vect X_1}=\hat U^\dagger_{t}\hat T_{\vect X_1}$ instead of $T_{\vect X_1}$).
This correspondence allows one to use two equivalent descriptions of the folded mini Brillouin zone (mBZ) of the electrons  in the presence of the moir\'e pattern, Fig.~\ref{fig:moire_BZ}(b). One, based on the longer periodicity implicit in the $ e^{i\vect\beta_m\cdot \vect r}$ dependence of the intervalley part of the Hamiltonian $\hat H$, suggests plotting the miniband dispersion over the smaller mBZ. The other, adjusted to the periodicity of the geometrical arrangement of atoms, uses the three times larger mBZ. 
For the smaller mBZ, the Dirac cones from both $K_+$ and $K_-$ valleys are folded onto the center of the mBZ, resulting in the valley degenerate dispersion surfaces shown in the left panel of Fig.~\ref{fig:bandstructure}(a). In contrast, the zone folding into the larger mBZ, shown in the center panel, places  Dirac cones from graphene's two valleys at opposite mBZ corners. 
The folding of dispersion surfaces from the larger mBZ into the smaller mBZ can be used relate the spectra shown in these alternative schemes. 
The unfolding of smaller mBZ into the larger mBZ is provided by the gauge transformation  $\vect \psi\rightarrow U\vect \psi'$, $\hat H\rightarrow \hat H'=U^\dagger \hat H U$ where $U=e^{\frac{i}{2}(\vect b_0+\tau_3\vect \beta_0)\cdot \vect r}$ represents a valley dependent shift of momentum.
After this gauge transformation, the new Hamiltonian $\hat H'$ can be written solely in terms of the $\check{\vect b}$ harmonics, 
 \begin{align}
\!&\hat H'  = v\left(\hat{\vect p}+\frac{1}{6}\left[3\vect b_0+\tau_3(\vect b_4+\vect b_5)\right]\right) \!\cdot  \vect \sigma\label{eq:H'}\\
&\qquad+\! U_{E'}vb\!\left(\tau_1\text{Re}  f' \!-\!\tau_2\text{Im} f' \right)\!\sigma_3 \nonumber\\
&\qquad+U_{\!G}v \! \left( \tau_1\text{Re} \,\vect g'\!-\!\tau_2\text{Im} \,\vect g' \right)\!\vect\sigma\nonumber\\
&\qquad+U_{\!G'}v \! \left( \tau_1\text{Re} \left[\hat R_{\frac{\pi}{2}}\vect g'\right]\!-\!\tau_2\text{Im}\left[\hat R_{\frac{\pi}{2}} \vect g' \right]\right)\!\vect\sigma
;\nonumber\\
&f' \!  =\! \frac{2}{\sqrt 3} \!\left(\! 1\! +\! e^{i\vect b_1\!\cdot\vect r} \!+\! e^{i\vect b_2\!\cdot\vect r} \!\right)\!,\,
 \vect g' \!  = \!\frac{2i}{\sqrt 3}\! \left( \!\vect b_0 \!+\! \vect b_2 e^{i \vect b_1\!\cdot \vect r} \! +\! \vect b_4e^{i\vect b_2\!\cdot\vect r} \right) \!.\nonumber
\end{align}

%%%%%%%%%%%%%%%%%% Fig 2 %%%%%%%%%%%%%%%%%%%

%%%%%%%%%%%%%%%%%% Fig 3 %%%%%%%%%%%%%%%%%%%
\begin{figure}[htbp]
  \includegraphics[width=0.48\textwidth]{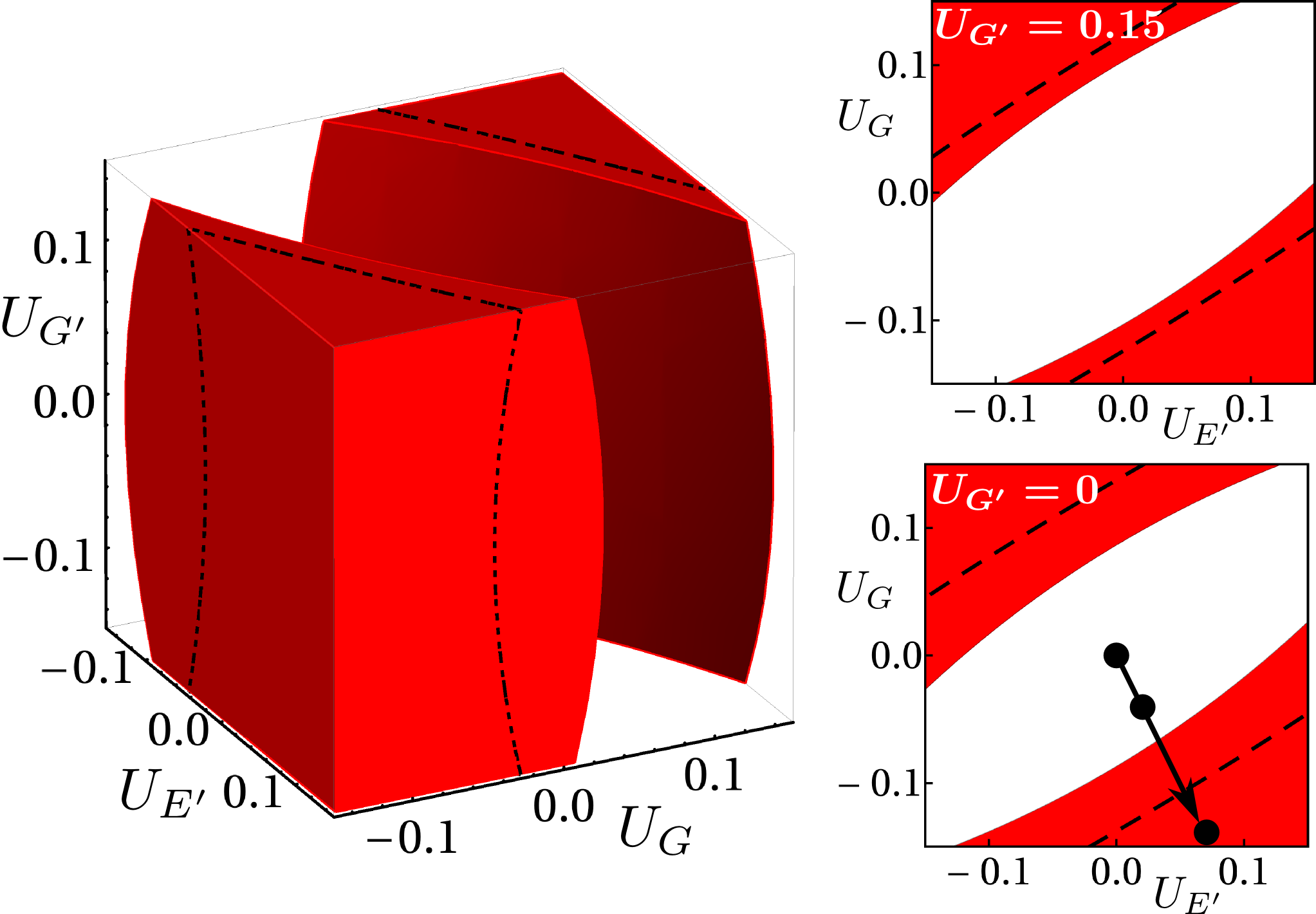}
  \caption{The regions of parameter space for which a band gap is present in the conduction band. The parameter space for the valence band is obtained by flipping the sign of $U_{E'}$.}
  \label{fig:para_plot}
 \end{figure}

%%%%%% Miniband spectra %%%%%%%%
Characteristic miniband spectra, calculated by numerical diagonalization in the basis of zone-folded plane waves of $K_+$ and $K_-$ Dirac electrons, are shown in Fig.~\ref{fig:bandstructure}(b,c).
The choices of phenomenological parameters used to calculate these spectra, marked with black dots in the lower right panel of Fig.~\ref{fig:para_plot}, correspond to the direction in the parameter space set by Eq.~\eqref{eq:model} with $\theta=0$.
Since nesting obscures some of the dispersion branches, it is useful to plot them over both the smaller mBZ (left) and the larger mBZ (middle).
Also, we note that, the calculated spectra will be electron-hole asymmetric, $\epsilon(\vect k)\neq-\epsilon(\vect k)$, unless either $U_E=0$ or  $U_G=U_{G'}=0$.

Generically, we find either a gapped edge of the first moir\'e miniband (on the conduction and/or valence band side of the graphene spectra) for a strong moir\'e perturbation, or gapless spectra with overlapping minibands for a weak moir\'e perturbation. In all cases, the main Dirac point is preserved with a renormalized Dirac velocity, $(1 - 12 U_{E'}^2- 24 U_{G}^2) v$. The parameter range where the spectrum has a gap at the first miniband edge in the conduction band is shown in red in Fig.~\ref{fig:para_plot}, whereas the parameter range with a gapless spectrum is left transparent.
The magnitude of the band gap between the first and second minibands in either the conduction band ($s=1$) or the valence band ($s=-1$), may be expressed in the form
\begin{align}
&\Delta=\frac{vb}{\sqrt 3}\min\left(c,d\right)\label{eq:band_gap}\\
&c\approx \!\frac{-1}{2}\!+\!|4U_{\!E'}\!-\!6s U_{\!G}| \!+\!\frac{4}{3}\!\left(\!U_{\!E'}^2 \!+\! 6 s U_{\!E'} U_{\!G}\! -\! 2 U_{\!G}^2\!-\!3U^2_{G'}\right)\nonumber\\
&d\approx|U_{E'}-2sU_G|+\frac{3}{2}\left(3 U_{E'}^2-4U^2_{G'}\right).\nonumber
\end{align}
where $\frac{vb}{\sqrt3}c$ and $\frac{vb}{\sqrt3}d$ are the values of the indirect and direct band gaps. A negative value of $\Delta$ indicates that the bands are overlapping (no band gap, transparent volume of Fig.~\ref{fig:para_plot}). 

%%%%%%%%%%%%%%%% Table 1%%%%%%%%%%%%%%%%%%%%%%%%%%%%
\begin{table}[t]
\caption{Almost commensurate $\sqrt3\times\sqrt3$ substrates for graphene.}
\begin{tabular}{l l l l }\hline
\textbf{Substrate} & $a_s$ \textbf{[\AA]} & \textbf{Structure} & \textbf{Ref}\\\hline\hline
$PtTe_2$  & 4.03  & layered & \cite{furuseth_acta_1965}\\\hline %checked with furuseth_acta_1965
$PdTe_2$ & 4.04 & layered & \cite{wilson_api_1969}\\\hline %checked with furuseth_acta_1965
% $PdTe_2$ & 4.04 & TMD tig prism, layered & \cite{furuseth_acta_1965,wilson_api_1969}\\\hline %checked wiht furuseth_acta_1965
$In Se$ & 4.05 & layered  & \cite{mccanny_ssp_1966} \\\hline %do better with this ref% NEED TO INC S.A. Semiletov Kristallogra.  3 288 (1958)
$h-GaTe$ & 4.01-4.06 &layered & \cite{zolyomi_prb_2013,semiletov_sovphys_1964,gillan_chemmater_1997}\\\hline%checked with gillan_chemmater_1997
$InP$ & 4.15 & (111) surface & \cite{Madelung_1991}\\ \hline %checked with Madelung_1991
$InAs$ &  4.28 & (111) surface & \cite{Madelung_1991}\\ \hline %checked with Madelung_1991
$GaSb$ &  4.31  & (111) surface & \cite{Madelung_1991}\\ \hline %checked with Madelung_1991
$AlSb$ &   4.34 & (111) surface & \cite{Madelung_1991}\\\hline\hline %checked with Madelung_1991
$\sqrt 3a$ graphene & 4.26 & &\cite{Castro_Neto_RevModPhys_2009}\\\hline
\end{tabular}
 \label{table_1}
\end{table}

%%%%%%% Conclusions %%%%%%%%%%%%%%%%%%
To conclude, there are numerous substrates with surfaces that are almost commensurate with the $\sqrt3\times\sqrt3$ Kekul\'e superlattice in graphene (Table \ref{table_1}). A sufficiently strong moir\'e perturbation for the Dirac electrons in graphene, placed at a small misalignment angle on such surfaces, results in a band gap between the first and second moir\'e minibands, on either the conduction or valence band side of graphene's band structure, at energies  $\epsilon\sim  \pm vb/\sqrt 3 $. These band gaps may be either indirect (if the phenomenological parameters $[U_{E'},U_G,U_{G'}]$ lie within the red volume of Fig.~\ref{fig:para_plot} between the black dashed lines), or direct (outside the black dashed lines). This observation suggests a new possibility to tailor electronic properties of graphene.

This work has been supported by EPSRC DTC NOWNANO, ERC Advanced Grant \emph{Graphene and Beyond}, Royal Society Wolfson Research Merit Award, and EPSRC Science and Innovation Award.


\begin{thebibliography}{99}
\bibitem{footnote:inversion}
Inversion asymmetric terms may be included into Eq.~\eqref{eq:H} by adding terms with $f_1(\check{\vect v})\rightarrow f_2(-\check{\vect v})$ and $f_2(\check{\vect v})\rightarrow f_1(-\check{\vect v})$.
\bibitem{footnote:model}
For the point charge model \cite{wallbank_prb_12},\\
$\tilde V=I\left( 4\pi/(3a)  \right)$, $\tilde v=I\left( 4\pi/(\sqrt3 a)\right)$,\\
$I(g)\approx\frac{8 \pi |\text e|Q}{ 9a^4\epsilon}\int\! dq_z dq'_z\frac{\psi^*(\vect K,q_z)e^{i(q_z-q'_z)\cdot d} \psi(\vect  K,q'_z)}{g^2+(q_z-q'_z)^2}.$\\
Here $Q$ is the charge per substrate lattice site, $D$ the graphene-substrate distance, $\epsilon$ the electric permittivity and $\psi(\vect Q,q_z)$ the Fourier transform of the graphene P$^z$ orbitals. 
For the hoping model $\tilde V= \tilde v= \frac{-\gamma^2}{9 \epsilon_s}$ with $\gamma$ the hopping integral to the substrate and $\epsilon_s$ is the energy of the substrate state.
\end{thebibliography}
\end{document}